\documentclass{svjour3} 

\smartqed

\newcommand{\be}{\begin{equation}}
\newcommand{\ee}{\end{equation}}
\def\lsim{\lower.5ex\hbox{$\; \buildrel < \over \sim \;$}}
\def\gsim{\lower.5ex\hbox{$\; \buildrel > \over \sim \;$}}

\usepackage{graphicx,amsmath}

\begin{document}
\title{Latest Trends in the Study of Accretion and Outflows Around Compact Objects}
\titlerunning{Accretion onto compact objects}
\author{Sandip K. Chakrabarti}
\authorrunning{S.K. Chakrabarti}

\institute{{S. N. Bose National Centre For Basic Sciences\\
	JD Block, Salt Lake, Sector-III, Calcutta-700091, India\\
	email: chakraba@boson.bose.res.in}}

\maketitle

\begin{abstract}
Study of astrophysics of black holes and neutron stars has taken a new
turn in the present decade with the realization that sub-Keplerian flows and the 
associated centrifugal barrier near the horizon or the surface 
of a neutron star play a major role in deciding the nature of 
the emitted spectra and the formation of outflows from the 
accreting matter. This region may remain steady or oscillate 
depending on the accretion rate, specific angular momentum 
and specific energy of the flow. Intricacies of oscillation 
may depend on the degree of feedback the inflow receives from the outflow. This region 
may emit hard or soft X-rays depending on relative numbers of hot elections
and soft photons intercepted by this region. We discuss how these 
properties come about and how they explain the observational results
of black hole candidates.

\end{abstract}

\noindent {\bf Published in Indian Journal of Physics, 1999, v. 73B, No. 6, p. 931-944}

\section{Introduction}

Several reviews on accretion processes on black holes and neutron stars 
have been written recently [1-3] where it is discussed that
sub-Keplerian flows play major role in explaining hard/soft state transition, 
quasi-periodic oscillation of X-Rays and outflows from accretion around black holes. 
In the present review, we shall concentrate on observational results
and how they could be explained with the latest solutions of accretion flows.

\section{Generic properties of the advective accretion flows}

Since the emergence of the standard Keplerian disk model [4] 
in seventies, purely rotating thick accretion disk models [5]
were built in the eighties, with nearly constant 
specific angular momentum $\lambda$, or where the angular momentum varies
with radial distance in the form of a power-law [6]. These are the predecessors
of current advective disk models [1-3], but the current models 
self-consistently include radial motions as well. Certain old models did attempt to
include  radial motion in the past [e.g., 7-6], but correct global solutions
were not found.

An accreting matter must have a significant radial motion, especially close to the
compact object. However, while on the horizon of a black hole, the velocity of matter 
attains the velocity of light, the matter must slow down and eventually stop on the hard surface
of a neutron star. This behaviour is completely independent of the outer 
boundary condition. For any generic equation of state, this means that matter 
must be supersonic on the horizon and subsonic on a neutron star surface. A supersonic
flow must be sub-Keplerian, and therefore the black hole accretion must deviate from 
a standard Keplerian disk [4]. Second, since infall time is very short 
compared with the viscous time to transport angular momentum, the specific angular
momentum ($\lambda$) must be almost constant. This means that the centrifugal force
$\lambda^2/r^3$ grows rapidly compared to the gravitational force $1/r^2$ and 
slows down the matter forcing it to be subsonic. Due to general 
relativistic effects (where gravity is really stronger than $1/r^2$ very close to the
horizon), matter recovers from this quasi-stagnant condition, 
passes through another sonic point and finally enters into the black hole supersonically.

This temporary stagnation of matter due to centrifugal force (at the so-called centrifugal 
force driven boundary layer of a black hole, or CENBOL), especially true when viscosity 
is low [9], at a few to a few tens of Schwarzschild radii is of 
extreme importance in black hole physics. Slowed down matter would be hotter 
and would emit harder radiations. If viscosity is high, the Keplerian disk extends (from outside)
to distances very close to the black hole, perhaps as close as the marginally stable orbit
($3R_g$, $R_g$ being the Schwarzschild radius.). If viscosity is low, the inner edge may
recede outwards to a few tens to a few hundreds of Schwarzschild radii. This is because
angular momentum must be transported outward by viscosity and smaller viscosity
transports slowly taking longer path length to reach a Keplerian disk.
When the viscosity is so high that the specific angular momentum rapidly varies with distance,
the centrifugal barrier weakens and the stagnation region disappears [10] since the 
sub-Keplerian region is confined roughly between $3R_g$ and $R_g$ only.

When the stagnation region forms {\it abruptly} in a fast moving sub-Keplerian flow,
a shock is said to have formed. Depending on the parameter space [11, 2-3],
this shock may or may not be standing at a given radius. It may form and propagate to
infinity [12], or may just oscillate [13-15], or may stand still [16-17]. In any 
case, generally shock forms and it decidedly affects the nature of the spectra
manifesting itself through propagating noise, or quasi-periodic oscillation [13-15], or 
steady state spectrum [18-19], or the formation of the quiescence state [20].

For a neutron star, the entire accretion may be sub-sonic 
and deviation from a Keplerian disk is not essential except in the narrow 
boundary layer where the rotational motion of the accreting matter
must adjust to the rotational motion of the star. This is true when magnetic field
is absent. In presence of a strong magnetic field, matter is stopped way ahead of the
stellar surface and is bent back along the field line. If the flow deviates from 
a Keplerian disk, the ram pressure $\rho v_r^2$ is higher and some matter can penetrate
radially through the field line and directly hit the surface when the field is
weaker. If this flow does become supersonic at any point, it must become subsonic 
just before hitting the surface [21].  

Formation of a shock or the stagnation region (boundary layer) is generic in 
neutron star accretions. This is because the flow must stop on the surface and a boundary
layer must be produced. However, in many cases, two shocks or stagnation
regions may form. Chakrabarti [11] showed (see also, [9])
that for a neutron star, one shock is possible at a large distance (say, $r\gsim10-15R_g$, 
where $R_g=2GM/c^2$ is the Schwarzschild radius, $G$, $M$ and $c$ are 
the gravitational constant, mass of the black hole and the
velocity of light respectively.) as in a black hole and 
another shock is possible at around $2.3-2.4 R_g$. Oscillations
of the outer shock would produce $1-10$Hz and the oscillation of the
shock at the inner shock would produce $\sim$KHz oscillations. However,
as Chakrabarti [9, 11] pointed out, inner shock would be possible only if the
star is compact enough, otherwise, it would a normal boundary layer (as 
in a white dwarf [22]), which could also oscillate if the cooling time matches with 
the infall time [13], or, if the input parameters are such that a steady 
solution is not possible as in a black hole accretion [14].

In the centrifugal barrier dominated boundary layer of black holes
and neutron stars, winds may form [2-3]. However, this phenomenon is more significant for a
black hole than for a neutron star. Neutron stars are known to have magnetic
fields and therefore most of the matter is stopped by it (unless the field is so weak that
sub-Keplerian matter with larger ram-pressure penetrates it easily and hits the
surface directly). Matter moves along the
field lines to the polar region and most of the winds come out from that region only.
On the contrary, black holes may not anchor magnetic fields. Most of the 
outflows can thus form at the CENBOL itself, due to its high outward pressure
gradient force, and due to the high temperature. However, it is unlikely that 
purely hydrodynamic acceleration could cause a `super-luminal' jet to form [23].
Since in the low-luminosity hard state CENBOL is hotter, the ratio $R_{\dot m} 
= \frac{{\dot M}_{out}}{{\dot M}_{in}}$ should be higher in hard states 
than in soft-states (when the CENBOL is cooler) [20].

There is another region of outflow: at the boundary between the
Keplerian and the sub-Keplerian matter. As Chakrabarti et al. [24-25] showed
through extensive numerical simulation, angular momentum distribution 
at this transition region is super-Keplerian and therefore, outward centrifugal 
force is very strong. Thus, a good deal of centrifugally driven outflow
at this region cannot be ruled out. In case of neutron stars, because of the
same reason, outflows are possible not only in polar caps, but at the transition
radius as well. If the magnetic field is non-aligned with the rotational axis, 
the Coriolis force on both sides of the disk would be slightly different and the 
QPO frequencies would split. Such behaviours have been observed in several 
neutron stars [26] and these essentially verify outflow solutions 
from transition regions [24-25].

\section{Spectral properties from compact objects: expectations vs. observations}

\subsection{Steady state properties}

If one denotes photon flux to be of the form, $\Phi_\nu \propto \nu^{-{\alpha+1}}$, 
(where $\nu$ is the photon frequency and $\alpha$ is the spectral index) then the energy spectrum would 
be $F_\nu =\int \Phi(\nu) d \nu \propto \nu^{-\alpha}$. The power is measured by
$P_\nu=\int F_\nu d\nu \propto \nu^{-{\alpha-1}}$. When the power is larger
in the hard X-ray region, the black hole is said to be in a hard state, whereas, when the
power is larger in the soft X-ray region, the black hole is said to be in a 
hard state. Several black hole candidates are known to switch between hard
and soft states [27-29].

In Fig. 1 (taken from [30], see also [9], [18]), a series of global solutions 
of the hydrodynamic equation with different viscosities are shown. The ratio $\lambda/\lambda_K$ 
(Here, $\lambda$ and $\lambda_K$ are the specific angular momenta of the disk and 
the Keplerian angular momentum respectively.) is plotted as a function of the 
logarithmic radial distance. The coefficients of viscosity 
parameter are marked on the curves (see, [30] for other parameters). 
At $x=x_{K}$, the ratio $\lambda/\lambda_K=1$ and therefore $x_{K}$ 
represents the transition region where the flow deviates 
from a Keplerian disk. First, note that when other parameters (basically, specific 
angular momentum and the location of the inner sonic point) remain roughly the same,
$x_{K}$ changes inversely with viscosity parameter $\alpha_\Pi$ [9]
(Only exception is the curve marked with $0.01$. This is because it is drawn
for $\gamma=5/3$; all other curves are for $\gamma=4/3$.). If one assumes [9, 18],
that alpha viscosity parameter {\it decreases} with vertical height, 
then it is clear from the general behaviour of Fig. 1 above that $x_K$ would 
go up with height from the equatorial plane. The disk will then look like a sandwich with higher 
viscosity Keplerian matter flowing along the equatorial plane. Soft photons coming 
out of the Keplerian component are intercepted by the sub-Keplerian region and are 
up-scattered (inverse Comptonization) to produce power-law hard X-rays. 
When the Keplerian disk rate is high,
there are enough soft photons to cool down the CENBOL completely 
and only soft X-rays are seen (except for a very weak power-law hard 
X-ray component extending upto a few hundred KeV about which we shall 
talk later). When the Keplerian rate is 
very small compared to the sub-Keplerian rate, the CENBOL cannot be cooled down,
and spectrum is dominated by hard X-rays only.  The Comptonization process
which is responsible is `thermal' Comptonization, since the energy received by
photons are coming from the thermal motion of the hot elections.

\begin{figure}
\includegraphics[width=0.8\textwidth]{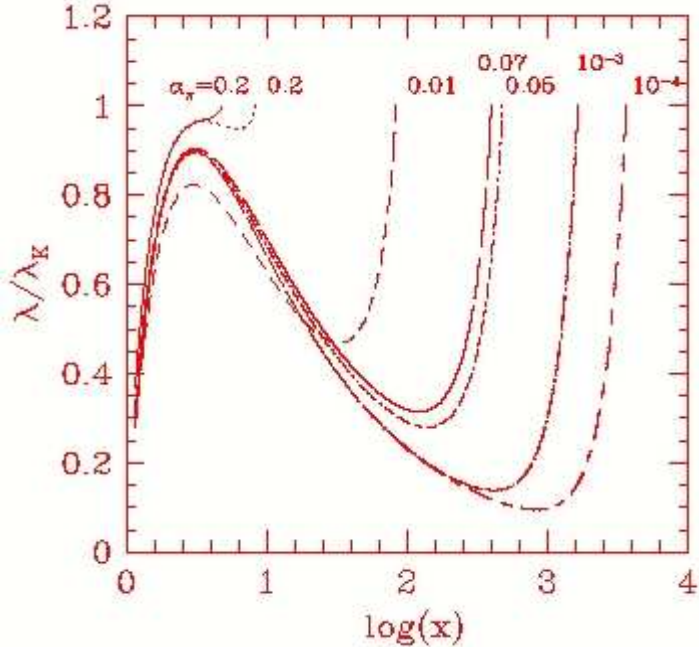}
\caption{Variation of $\lambda/\lambda_K$ with logarithmic radial distance for
a few solutions. Viscosity parameter $\alpha_\Pi$ is marked on each curve. 
Note that except for the dashed curve marked $0.01$ (which is for $\gamma=5/3$, and the rest are for
$\gamma=4/3$), $x_K$, where $\lambda/\lambda_K=1$, generally 
rises with decreasing $\alpha_\Pi$. Thus, low viscosity
flows must deviate from the Keplerian disk farther out of the black hole [30]. }
\end{figure}

Independent of the outer boundary conditions (accretion rate, viscosity etc.),
since matter must rush towards the black hole with a high velocity (after crossing
the inner sonic point), a new phenomenon is observed as
first noticed by Chakrabarti \& Titarchuk [18]. This is known
as  the bulk motion Comptonization. Radiation is up-scattered as it hits 
radially moving electron, independent of whether the electrons themselves are
hot or cold. In the hard state, this effect of up-scattering due to bulk motion is washed out
by the  up-scattering due to thermal motion. But in the soft state, when the 
electrons are cooled down to a few KeVs [18], this particular kind of up-scattering
produces a power-law tail extending to a few hundred KeV. This is the true signature
of a black hole accretion. A neutron star accretion would not have the bulk motion
Comptonization and the power-law tail is not observed either. Several black holes
have now been identified using this technique [31-32]. This method is clearly useful
when one considers massive and supermassive black holes [33].

As the viscosity changes at the outer edge, the sub-Keplerian and Keplerian 
flows redistribute [12] and the inner edge of the Keplerian component 
also recedes or advances. When the Keplerian component advances (high viscosity,
high rate), the soft state is achieved. When the Keplerian component
recedes, the hard state is achieved. This behaviour is seen in many black hole
candidates [32, 34].

On the other hand, to change states (hard to soft and vice versa), it is sufficient 
if the sub-Keplerian component alone changes [18]. Yadav et al. [35] pointed 
out that the state change takes place in GRS1915+105 in matter of a few tens of 
seconds during quasi periodic oscillations. Since the viscous time-scale is much 
longer, the inner-edge of the Keplerian flow does not have time to move in or out within
these short periods. Hence, appearance and disappearance of sub-Keplerian 
flows must be responsible. The exact mechanism will be discussed in detail later. 

\subsection{Time dependent properties}

After years of X-ray bursts, an X-ray novae can become very faint
and hardly detectable in X-rays. This is called the
quiescent state. This property is built into the advective disk models. As already 
demonstrated (Fig. 1) $x_{K}$ recedes from the black 
holes as viscosity is decreased. With the decrease of
viscosity, less matter goes to the Keplerian component i.e., ${\dot m}_d$ goes down. Since the 
inner edge of the Keplerian disk does not go all the way to the
last stable orbit, optical radiation is weaker in comparison
with what it would have been predicted by a standard disk model [4]. 
This behaviour is seen in V404 Cyg [36] and A0620-00 [37].
The deviated component from the Keplerian disk almost resembles a constant 
energy thick ion torus of earlier days [38]. 

In some large region of the parameter space
the solutions of the governing equations are inherently
time-dependent [1-3]. Just as a pendulum inherently oscillates while searching
for a static solution, the physical quantities of the advective disks also show 
oscillations of the CENBOL region for some range in parameter space. This 
oscillation is triggered by competitions among various time scales (such as infall 
time scale, cooling time scales by different processes, including matter loss through 
outflows). Thus, even if black holes do not have hard surfaces, quasi-periodic oscillations could
be produced. Although any number of physical processes such as
acoustic oscillations [39], disko-seismology [40], or trapped oscillations [41]
could produce such oscillation frequencies, modulation of $10-100$ 
per cent or above cannot be achieved without bringing in the dynamical 
participation of the hard X-ray emitting region, namely, the CENBOL.
By expanding back and forth (and puffing up and collapsing, alternatively)
CENBOL intercepts variable amount of soft photons
and reprocesses them. A typical observation is presented by Rao [42, see also, 43-44]. 
These observations can be readily explained by combination of cooling processes in CENBOL.

Imagine that an otherwise standing shock wave is perturbed to move outward against 
the direction of the inflow. The post-shock flow would be hotter than 
the steady-state value, and the cooling rate would be higher. The increased 
cooling would reduce the post-shock pressure which will cause collapse of the shock
towards the black hole. In this case, the post-shock temperature would be lower,
and the cooling will slow down. Ram pressure and thermal pressure of the incoming flow
would push the shock inward where the post shock-pressure is higher. This causes
overshooting of the shock location and the shock starts moving outwards. This 
oscillation is quasi-sinusoidal in presence of a single cooling process, such 
as bremsstrahlung and Comptonization [13], but could be more complex looking if 
a large number of cooling processes are included. A outflow from CENBOL is also 
a kind of dynamical `cooling' and the presence of an outflow causes evolution of the
pattern of the oscillation such as shown in [42-44] for the galactic black hole
GRS 1915+105. Details of the effects of the outflow may be in order in this context.

\subsection{Effects of outflows on quasi-periodic oscillations}

Chakrabarti \& Bhaskaran [45] first pointed out that outflows are
more easily produced from a `sub-Keplerian' flow, {\it even if}
they are centrifugal pressure driven. A simple computation of the outflow
from CENBOL area based on quasi-spherical accretion yields [2-3, 46] the ratio
of outflow to inflow rate to be:
$$
\frac{{\dot M}_{out}}{{\dot M}_{in}} =R_{\dot m}=
\frac{\Theta_{out}}{\Theta_{in}}\frac{R}{4} 
[f_0]^{3/2} exp  (\frac{3}{2} - f_0)
\eqno{(1)}
$$
where, $R$ is the compression ratio of the flow at the CENBOL 
and $\Theta_{in}$ and $\Theta_{out}$ are the 
solid angles of the inflow and the outflow respectively. $f_0=R^2/R-1$.
Location of the sonic point of the outflow is at $r_c=f_0r_s$ [3, 46].

Figure 2 shows the ratio $R_{\dot m}$ as a function of the compression ratio 
$R$ (plotted from $1$ to $7$) assuming $\Theta_{out} \sim \Theta_{in}$ 
for simplicity. Note that if the compression (over and above the
geometric compression) does not take place (namely, if the CENBOL
does not form), then there is no outflow in this model. Indeed, for $R=1$,
the ratio $R_{\dot m}$ is zero! Thus the driving
force of the outflow is primarily coming from the hot, compressed region.
This picture is qualitatively supported by more detailed analysis [20].

\begin{figure}
\includegraphics[width=0.8\textwidth]{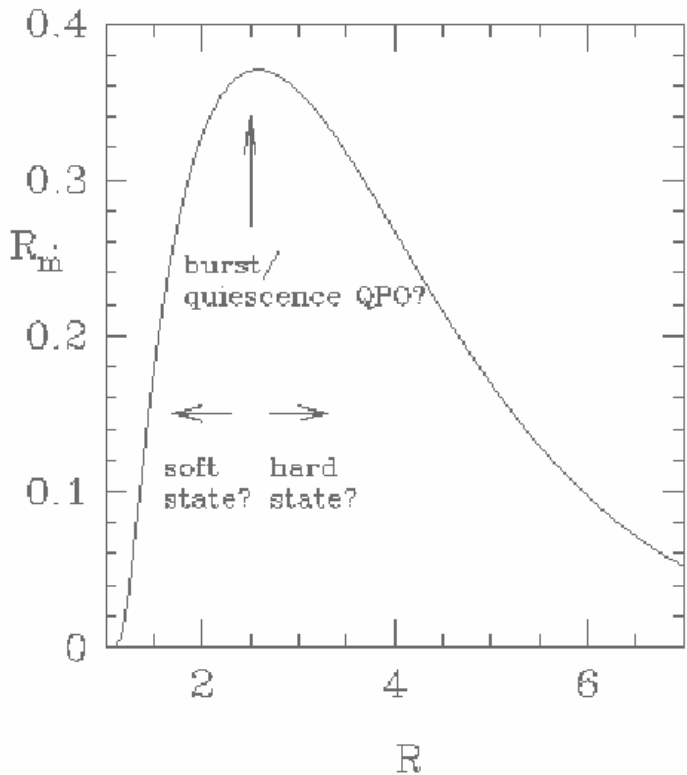}
\caption{Ratio ${\dot R}_{\dot m}$ of the outflow rate and the inflow rate
as a function of the compression ratio of the gas at the dense region boundary.
Solid angles subtended by the inflow and the outflow are assumed
to be comparable for simplicity. Note that the ratio peaks at intermediate compression $R\sim 2.5$.
Burst/quiescence feature of transient sources are related to this behaviour of the outflow rate.}
\end{figure}

One can discuss the effects of outflows [46] of a stellar black candidate such as GRS 1915+105
which exhibits QPOs as well as periodic changes to burst and quiescences state in a matter of
tens of seconds. Particularly interesting is that it shows QPO of around $\nu_I=1-10$Hz and 
sometimes bursts at around $\nu_H=0.01$ Hz [43]. It also seems to show a 
QPO at $\nu_H\sim 67$Hz. It has been pointed out before that the spectrum of CENBOL may be softer
in presence of outflows [47]. This is because CENBOL would have lesser matter (for a given
soft photon flux). If the behaviour of outflow rate truly depends on compression ratio
as in Fig. 2, one may be able to explain many complex properties, though details have to be
worked out. An interesting property of shock-compressed $R_{\dot m}$ (Fig. 2) is that 
the outflow rate is peaked at around $R=2.5$, i.e., when the
shock is of `average' strength. On either side, the outflow rate falls off very rapidly.
As $R_{\dot m}$ [Eq. (1)] is independent of shock location $r_s$, shocks oscillating with time period
similar to the infall time in $r<r_s$ region (causing a 1-10Hz QPO 
in the hard state [13, 18-19]) will continue to have outflows and gradually fill the 
(sonic) sphere of radius $r=r_c=f_0 r_s/2$ ($f_0=R^2/R-1$) (where matter is slowly moving) till
$<\rho> r_c \sigma_T \sim 1$ when this region would be cooled down
catastrophically by inverse Comptonization. At this stage: (a) $r<r_c$ region 
would be drained to the hole in $t_{fall} \sim r_c^{3/2} 2GM/c^2 = 0.1 (\frac{r_s}{50})^{3/2} 
(\frac{M}{10 M_{\odot}}) (\frac{f_0}{4})$ seconds. This is typically what is observed 
for GRS1915+105. [35]. If on the other hand, the outflow becomes locally supersonic
(due to sudden cooling) the flow would run away outward forming a shock in the outflow. (b) In a cooler
accreting gas, the shock will disappear, and a smaller compression ratio 
($R\rightarrow 1$) would `cut off' the outflow (Fig. 2).
In other words, during burst/quiescence QPO phase, the outflow would be blobby. 
(c) The black hole will go to a soft state during a short period. 
If the angular momentum is large enough, this brief period of 
soft state may be prolonged to a longer period of tens of seconds depending on how
the centrifugal barrier is removed by viscosity (generated during shock oscillations).
There are actually two possibilities, if the sonic sphere is cooled down, the outflow
could become locally supersonic, and the burst state becomes momentary. If on the
other hand, the flow in sonic sphere remains subsonic even after cooling, and returns back to the black hole
due to lack of driving force,
then the burst state may be long-lived when angular momentum is high, otherwise 
short-lived as discussed above. It is also possible that cooler outflow takes a 
longer time to be drained out, and caused prolonged burst state. A signature
of bulk motion Comptonization in the burst-state spectrum would ensure the drainage by the black hole.
Since $f_0\sim 4$, the sonic sphere is typically four times bigger and would intercept four times
larger number of photons. Thus the count rate should fluctuate roughly by a factor of $f_0$ in between
quiescence and burst states. This is precisely what is observed [48].

There is one good test to check if the QPOs are actually from shock oscillations.
When CENBOL oscillates, the fractional change in hard X-ray emission is very high compared to the
fractional change in softer X-rays (since most of the contribution to soft X-rays 
is from the Keplerian disk whose properties are not modulated with QPOs). This is 
clearly demonstrated in this object (Fig. 5 of [48]). Similarly, when the
Keplerian accretion rate is increased, the cooling time goes down, and one would expect the 
frequency to increase with soft luminosity. In soft states QPOs are usually absent, but in 
burst phase if the QPO is present it is seen in higher frequency only (Fig. 2 of [48]).

Typically, a shock located at $r_s$ (measured hereafter in units of $R_g=2GM/c^2$),
produces an oscillation of frequency [13],
$$
\nu_{I} = \frac{1}{t_{ff}} \sim \frac{1}{R} r_s^{-3/2} \frac{c}{R_g}
\eqno{(2)}
$$
where, $R$ is the compression ratio of the gas at the shock
which the infall velocity goes down and thefore the infall
time $t_{ff}$ goes up. For a gas of $\gamma=4/3$, $R \sim 7$ and for $\gamma=5/3$,
$R\sim 4$ respectively when the shock is strong. Thus, for instance, for a
$\nu_I=6$Hz, $r_s \sim 38$ for $M=10M_\odot$ and $\gamma=4/3$. For
$\nu_H=67$Hz, $r_s \sim 8$ for the same parameters. Chakrabarti
\& Titarchuk [18] pointed out that since black hole QPOs
show a large amount of photon flux variation, they can not be explained
simply by assuming some inhomogeinities, or purturbations in the flow.

Since the sonic point location is a function of the shock location, one would 
expect that by time of filling the sonic sphere (and time to cool it, i.e., the 
duration of the quiescence state) would be correlated with the QPO frequency.
The volume filling time of the sonic sphere is [46],
$$
t_{fill}=\frac{4\pi r_c^3 <\rho>}{3 {\dot M}_{out}}
\eqno{(3)}
$$
The cooling starts when $<\rho>r_c \sigma_T \gsim 1$,  $\sigma_T=0.4$
is the Thomson scattering cross-section. Thus the duration of
the quiescence state is given by,
$$
t_{q} =  \frac{4\pi r_c^2}{3 {\dot M}_{out} \sigma_T}.
\eqno{(4)}
$$
Because of uncertainties in $\Theta_{in}$, $\Theta_{out}$ and ${\dot M}_{in}$ (see, eq. 1)
(subscript `in' refers to the inflow rate) we define a new parameter,
$$
\Theta_{\dot m} = \frac{\Theta_{out}}{\Theta_{in}} \frac{{\dot M}_{in} }{{\dot M}_{Edd}} .
\eqno{(5)}
$$
Using these equations, we get the expression for $t_q$ as,
$$
t_q=\frac{41.88} {(R-1)^{1/2}} \frac{r_s^2 R_g^2 exp(f_0-\frac{3}{2})} {{\dot M}_{Edd} \Theta_{\dot m}} \ \ s .
\eqno{(6)}
$$
(where we brought back $R_g$ factor to get cgs unit.)
Or, eliminating shock location using eq. (2), we obtain,
$$
t_q= 56.4 \frac{exp(f_0-\frac{3}{2})} {R^{4/3} (R-1)^{1/2}\Theta_{\dot m}} (\frac{M}{10 M_\odot})^{-1/3} \nu_I^{-4/3} s
\eqno{(7)}
$$
For an average shock, $2.5<R<3.$, the result is insensitive to
the compression ratio. Using average value of $R=2.9$ and
for $\Theta_{\dot m} \sim 0.4$ (which corresponds to $0.4$ Eddington
rate for $\Theta_{out} \sim \Theta_{in}$) we get,
$$
t_q= 461.5 (\frac{0.4}{\Theta_{\dot m}}) (\frac{M}{10M_\odot})^{-1/3} \nu_I^{-4/3} s
\eqno{(8)}
$$
Thus, the duration of the off-state must go down rapidly
as the QPO frequency increases if the net accretion remains fixed.
Similarly, if the hot post-shock gas of height
$\sim r_s$ intercepts $n$ soft photons per second,
from the pre-shock Keplerian component [18],
it should intercept about $f_0 n$ soft photons per second
when the sonic sphere of size $r_c$ is filled in. Thus, the
photon flux in the burst phase should be about $f_0=R^2(R-1)^{-1} \gsim  4$
times larger compared to the photon flux in quiescence state. Depending
on the degree of flaring, the interception may be much higher.

In Fig. 3 we show results of three observations of X-ray transient GRS 1915+105 (see, [48] for the observational 
results) where we plot $\nu_{I}$ along y-axis and $t_{q}$ along x-axis. 
The dotted curve shows eq. (8) which passes close to June 18th, 1997  data and Oct 7th, 1996. A short dashed
curve is drawn using the same equation but assuming a law $\nu^{-2}$ and for 
accretion parameter $\Theta_{\dot m}=0.15$. Several reasons, such as
slower infall velocity in presence of angular momentum and/or assumption constant volume filling
rate ${\dot M}_{out}/<\rho>$ would modify the eq. (8) to have an inverse square dependence on $\nu$.
However exact derivation with these new physical inputs would require more parameters such as
angular moementum or mean density.
The result of May 26th, 1997 falls on a different inverse square curve 
(long dashed) valid for $\Theta_{\dot m}=0.055$ indicating a change in
outflow geometry and/or a change in accretion rate.

\begin{figure}
\includegraphics[width=0.8\textwidth]{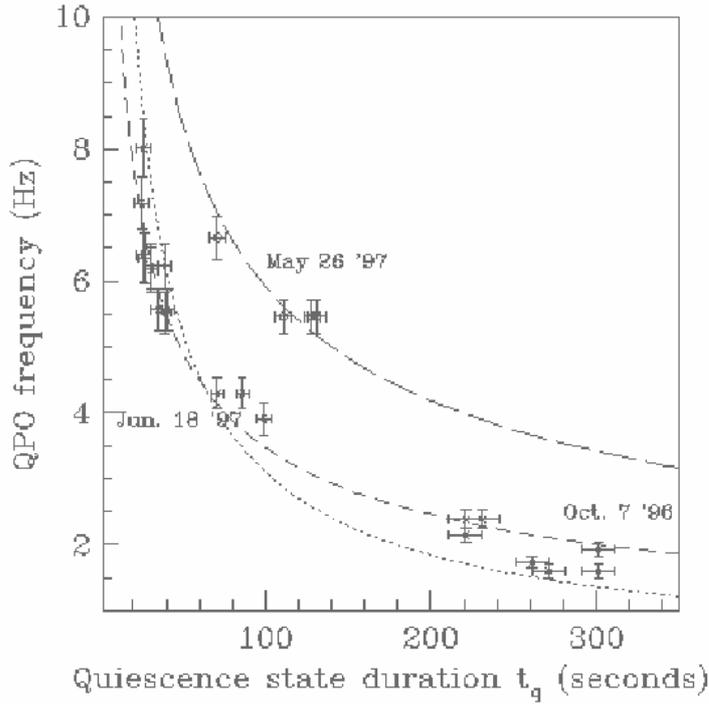}
\caption{Variation of QPO frequency $\nu_I$ with duration of quiescence
states $t_{q}$. Dotted curve is the $t_q \propto \nu^{-4/3}$ law (eq. 8) derived using simple
free-fall velocity assumption. Dashed curves are $t_q\propto \nu^{-2}$ fits
suggesting modified velocity law. The general agreement points to the
shock oscillation model}
\end{figure}

\section{Future directions}

Presence of a sub-Keplerian disk (hot or cold) close to a black hole has 
forced people to re-analyze some of the old problems and several new findings have
been made. For instance, recently it has been noted that significant nucleosynthesis 
inside this region can destabilize a disk and cause sonic point oscillations [30]. 
However, an old problem of in-spiraling and coalescence of two point masses 
have re-surfaced. It is recognized that the presence of a sub-Keplerian disk [49]
can seriously change the nature of gravitational wave emission if the accretion rate
is high enough. Because of its importance in next generation gravitational wave
astronomy, it is worth repeating some of the arguments.

Chakrabarti [50] first pointed out that the accretion disks close to the
black hole need not be Keplerian and it would affect the
gravitational wave properties. The radiation pressure
dominated disks are likely to be super-Keplerian in some range which would
transfer angular momentum to the orbiting companion and in some
extreme situations, can even stabilize its orbit from coalescing
any further. This was later verified by time-dependent numerical simulations [51].
Fig. 1 shows, however, that in most of the regions flow likes to 
remain sub-Keplerian ($\lambda/\lambda_K <1$) rather than 
super-Keplerian ($\lambda/\lambda_K >1$).
Assume that a companion of mass $M_2$ is in an instantaneous
circular Keplerian orbit around a central black hole of mass $M_1$.
This assumption is justified, specially when the orbital radius is
larger than a few Schwarzschild radius where the energy loss per orbit
is negligible compared to the binding energy of the orbit.
The rate of loss of energy $dE/dt$ in this binary system with orbital
period $P$ (in hours) is given by (see, [49] and references therein),
$$
\frac{dE}{dt}=3 \times 10^{33} (\frac {\mu}{M_\odot})^2
(\frac{M_{tot}}{M_\odot})^{4/3} (\frac{P}{1 hr})^{-{10}/{3}} {\rm ergs\ sec^{-1}},
\eqno{(4)}
$$
where,
$$
\mu=\frac {M_1 M_2}{M_1+M_2}
$$
and
$$
M_{tot}=M_1+M_2.
$$
The orbital angular momentum loss rate would be,
$$
R_{gw}=\frac{dL}{dt}|_{gw}=\frac{1}{\Omega} \frac{dE}{dt},
\eqno{(5)}
$$
where, $\Omega=\sqrt{G M_1/r^3}$ is the Keplerian angular velocity of the
secondary black hole with mean orbiting radius $r$. The subscript `gw'
signifies that the rate is due to gravitational wave emission.
In presence of an accretion disk co-planer with the
orbiting companion, matter from the disk (with local
specific angular momentum $\lambda(r)$) will be accreted onto the companion
at a rate close to its Bondi accretion rate [49],
$$
{\dot M}_2=\frac{4\pi {\bar \Lambda} \rho (GM_2)^2}{(v_{rel}^2+a^2)^{3/2}}
\eqno{(6)}
$$
where, $\rho$ is the density of disk matter, ${\bar \Lambda}$ is
a constant of order unity (which we choose to be $1/2$ for the rest
of the paper), $v_{rel}=v_{disk}-v_{Kep}$ is the relative
velocity of matter between the disk and the orbiting companion.
The rate at which angular momentum of the companion will be changed
due to Bondi accretion will be, 
$$
R_{disk}=\frac{dL}{dt}|_{disk}={\dot M_2} (\lambda_{K} (r) -\lambda (r) ) .
\eqno{(7)}
$$
Here, $\lambda_{K}$ and $\lambda$ are the local Keplerian and disk
angular momenta respectively. The subscript in the left hand
side signifies the effect is due to the disk. If some region 
of the disk is sub-Keplerian ($\lambda<\lambda_{K}$), the
effect of the disk would be to reduce the angular momentum of the
companion further and hasten coalescence. If some region of the
disk is super-Keplerian, the companion will gain angular momentum
due to accretion, and the coalescence is slowed down. 

In order to appreciate the effect due to intervention of the
disk, we consider a special case where, $M_2 <<M_1$, $\lambda<<\lambda_{K}$
and $v<<a$ (subsonic flow). In this case,
$\mu \sim M_2$ and $M_{tot}\sim M_1$. The ratio $R_{d-g}$ of these two rates is,
$$
R_{d-g}=\frac{R_{disk}}{R_{gw}}=1.518\times 10^{-7} \frac{\rho_{10}}{{T_{10}}^{3/2}}
{x^4}{M_8}^2
\eqno{(8)}
$$
Here, $x$ is the companion orbit radius
in units of the Schwarzschild radius of the primary,
$M_8$ is in units of $10^8 M_\odot$, $\rho_{10}$ is the density in units
of $10^{-10}$ gm cm$^{-3}$ and $T_{10}$ is the temperature of the
disk in units of $10^{10}$K. It is clear that, for instance, at $x=10$,
and $M_8=10$, the ratio $R\sim 0.15$ suggesting the effect of the
disk could be a significant correction term to the general relativistic
loss of angular momentum. The ratio $R_{d-g}$ is independent
of the mass of the companion black hole, as long as $M_2 <<M_1$.  
If the flow is highly supersonic ($v>>a$) then
the result is independent of temperature of the flow. 

Chakrabarti [49] shows that the effect is especially important 
when the net accretion rate (Keplerian and sub-Keplerian) is high enough. (Effects
shown in [49] were computed for ${\dot M}=1000{\dot M}_{Edd}$, for instance.
Clearly, the net effect goes down with accretion rate.) In a hard state,
although the Keplerian rate is low, the sub-Keplerian rate could be high and 
this effect is important. However, in the soft state, the disk remains 
basically Keplerian except last two-three Schwarzschild radii. Thus, the effect would 
be negligible except in this region. A combination of results from electromagnetic spectrum
and gravitational wave spectrum would be more effective to put limits on the
parameters such as the masses of the black holes, individual accretion rates and
separation between the black holes etc. Clearly this is going to be an exciting task.

An effect so far ignored in this context is the formation of outflows
from the CENBOL region [2, 3, 20, 52] which may `mess-up' computation
even farther. It has been shown [20, 51] that profuse outflows, to the extent causing
evacuation of the disk can be produced. In this case, the effect would clearly be
negligible, but the non-linearity of the outflow rate with inflow rate (Fig. 2) 
makes the effect more difficult to incorporate. 

\vspace{0.3cm}

\noindent {\bf Acknowledgments}\\

\noindent This work is partly supported by a grant from DST for the project `Analytical and numerical studies of
astrophysical flows around black holes and neutron stars'.

{}
\end{document}